\documentclass[iicol, pdflatex, sn-basic]{sn-jnl}

\usepackage{graphicx}%
\usepackage{subcaption}
\usepackage{multirow}%
\usepackage{amsmath,amssymb,amsfonts}%
\usepackage{amsthm}%
\usepackage{mathrsfs}%
\usepackage[title]{appendix}%
\usepackage[dvipsnames]{xcolor}%
\usepackage{textcomp}%
\usepackage{manyfoot}%
\usepackage{booktabs}%
\usepackage{algorithm}%
\usepackage{algorithmicx}%
\usepackage{algpseudocode}%
\usepackage{listings}%
\usepackage{siunitx}%
\usepackage{lmodern}
\usepackage{natbib}
\usepackage{soul}
\usepackage{hyperref}
\DeclareSIUnit\erg{erg}

\begin{document}

\title{Editorial: Mass and Angular Momentum Transport of Rapidly Rotating Hot Stars}

\author*[1]{\fnm{Paul A.} \sur{Scowen}}\email{paul.a.scowen@nasa.gov}

\author[2]{\fnm{Carol E.}
\sur{Jones}}\email{cejones@uwo.ca}
\equalcont{These authors contributed equally to this work.}

\author[3,4]{\fnm{Ren\'e D.} \sur{Oudmaijer}}\email{ r.d.oudmaijer@leeds.ac.uk}
\equalcont{These authors contributed equally to this work.}
\author[5]{\fnm{Jamie} \sur{Lomax}}\email{lomaxj@usna.edu}
\equalcont{These authors contributed equally to this work.}

\author[6,7]{\fnm{Jeremy J.} \sur{Drake}}\email{jeremy.1.drake@lmco.com}
\equalcont{These authors contributed equally to this work.}

\affil*[1]
{\orgdiv{NASA Goddard Space Flight Center}, 
{\orgname{Code 667}}, \orgaddress{\street{8800 Greenbelt Road}, \postcode{20771}, \state{MD}, \country{USA}}}

\affil[2]
{\orgdiv{Physics and Astronomy}, 
{\orgname{The University of Western Ontario}}, \orgaddress{\street{1151 Richmond Street}, \city{London}, \postcode{N6A 3K7}, \state{Ontario}, \country{Canada}}}

\affil[3]
 {\orgname{Royal Observatory of Belgium}, \orgaddress{\street{Ringlaan 3}, \city{1180 Brussels}, \country{Belgium}}}

\affil[4]
{\orgdiv{School of Physics and Astronomy}, 
{\orgname{University of Leeds}}, 
\orgaddress{\city{Leeds LS2 9JT}, \country{United Kingdom}}}

\affil[5]
{\orgdiv{Physics Department}, 
{\orgname{United States Naval Academy}}, \orgaddress{\street{572C Holloway Road}, \city{Annapolis}, \state{MD} \postcode{21402}}}

\affil[6]
{\orgdiv{Advanced Technology Center}, \orgname{Lockheed Martin}, \orgaddress{\street{3251 Hanover Street}, \city{Palo Alto}, \state{CA} \postcode{94304}}}

\affil[7]
{\orgname{Lockheed Martin Solar and Astrophysics Laboratory}, \orgaddress{\street{3251 Hanover Street}, \city{Palo Alto}, \state{CA} \postcode{94304}}}



\maketitle


Massive stars are the rarest, shortest-lived, yet most important contributors to galactic cosmic evolution (e.g. \citealt{Eldridge2022}).  They live out their entire lives and go supernova while low-mass stars are still forming, and drive the ecology of star formation through the Baryonic Cycle and ionizing radiation feedback \citep{Peroux2020}.   Moreover, the light from massive stars combined with population synthesis models is used to understand star-formation rates in galaxies at all distances \citep{Conroy2013}.

A substantial fraction of high mass stars are born as rapid rotators, with a larger fraction as moderate rotators where evolution is predicted to diverge from slowly rotating stars \citep[e.g.][]{deMink2013}. The effects of extreme rotation can drastically alter the evolution of the star thereby influencing the rate and nature of the resulting supernovae and the remnants they leave behind.

At this time, theories predict profound, yet quite different, scenarios and consequences for rapid rotation \citep[e.g.,][]{Voss2009}, involving binary mass transfer \citep[e.g.][]{deMink2013} and chemical mixing. Observational constraints are needed to resolve the many outstanding problems.  Ultraviolet (UV) spectropolarimetry provides a unique opportunity to exploit the stellar and wind aspherical geometries due to rapid rotation to constrain the internal physics that dictates the evolution of the star, and its impact on the Galaxy \citep{Ignace2024}. 

Viewed globally, massive stars are responsible for contributing the heavier elements --- the metals--- to the galactic ecology, coupled with considerable energy and ionizing radiation \citep{Leitherer1992, Schaerer1997}.  As such, the understanding of their evolution over their lives is a critical piece in understanding the evolution of galaxies and the production of material necessary for the formation of planetary systems and ultimately the conditions for life.  In comparison, our Sun is a low mass, slowly-rotating, single star. This means that many of the processes that we seek to understand about the role of rotation in affecting the evolution of a star can only be done by observation, modeling and analysis of massive stellar systems at distance, but still close enough to address them as individual systems with high enough photon counts for sufficient signal-to-noise.

\begin{figure*}
    \centering
    \includegraphics[scale=0.75]    
    {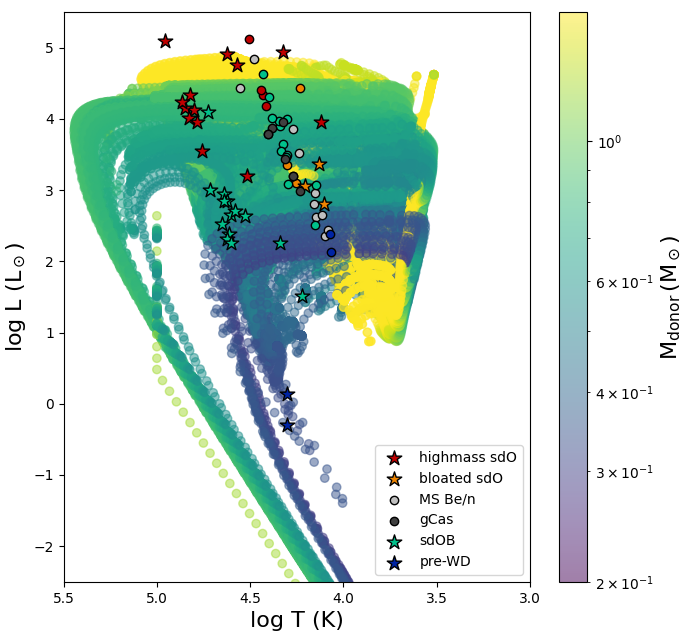}
    \caption{An HR diagram from Labadie-Bartz et al. (2025, this collection) showing evolutionary tracks extracted from a subset of 318 models using the Binary Population and Spectral Synthesis (BPASS) code \citep{Eldridge2017} of both the mass donor and mass gainer.}
    \label{fig:Fig1}
\end{figure*}

About one-fifth of massive stars are observed to rotate rapidly (projected rotational velocities $>$ 200 km/s, \citealt{deMink2013}), and they are often affected by binary interaction since more than 70\% of massive stars occur in binary pairs \citep{Sana2012, deMink2013}. The missing links in our knowledge boil down to (1) how and when rapid stellar rotation occurs, (2) what it does to the star and its environment, and (3) how the angular momentum and chemical enrichment are transported internally, with attention to the fact that stars must ingest an enormous amount of angular momentum in order to accrete mass from an orbiting companion. 

This dedicated journal collection will present and discuss a variety of science cases that can be used to extend our knowledge of massive stars and the influence of their rapid rotation on their subsequent evolution.  The aim is to build understanding of this pivotal class of stellar objects that provides the energy and processed material driving  galactic evolution and setting the stage for subsequent star and planet formation.

\begin{itemize}
\item{\bf Binary Be-stars} \\
See \citet{lab25}.

\smallskip
Massive stars come in a variety of flavors, from the most massive O-type stars down to the smaller but more numerous and equally influential B-stars.  There is a growing body of evidence that many B-stars that are seen to rapidly rotate were formed through the process of mass transfer in binary systems, and are now seen as parts of binary systems comprising mid- to late-type stars with a stripped core companion. Figure~\ref{fig:Fig1} shows evolution tracks for binary systems that are likely to undergo mass transfer resulting in a spun up B-type star. See Section~4.2 of the paper linked to this list item for more detail.
There is an opportunity now to locate and study the stripped cores of the mass donors in such systems using ultraviolet spectroscopic measurements to reveal intimate details about how such systems evolve and ultimately what the fate of such pairs of stars might look like.\\

\item{\bf Observational effects of rapid rotation for Be stars} \\
See \citet{ras25a}.

\smallskip
Progress in understanding Be stars has been stymied by the limitations of traditional observational approaches and the lack of instrumentation in a key spectral band: the far ultraviolet. Going further, there are a variety of modeling approaches that can be used to infer observable properties of rapidly rotating Be stellar systems across their ranges of spectral types using ultraviolet polarimetry combined with photometric colors and H$\alpha$ line profiles.  The effects of rapid rotation can intimately affect the measured equivalent width of the H$\alpha$ line and can provide a method to disentangle the effects of inclination, especially when gravity darkening from the resulting circumstellar disk is also included.  Crucially, the polarimetric signal measured from the star can also provide a sensitive measure of the actual rotation rate. For many Be stars, such measurements are only feasible in the ultraviolet where the intrinsic polarization signature is strongest.\\

\item{\bf Accreting companions of Be/$\gamma$ Cas stars}\\
See \citet{ras25b}.

\smallskip
One particular class of Be stars is characterized by bright X-ray emission and its members are named  after their prototype, $\gamma$ Cas.  There is strong suspicion that the energetic emission observed originates from mass transfer onto a white dwarf companion---dynamics that can be modeled using 3D Smoothed-particle hydrodynamical simulations. Figure~\ref{fig:Fig2} is a rendition of figure~1, section~3, in the paper linked to this list item. Notice that after the quasi-steady state is reached, as shown, that an accretion disk has formed around the secondary star. Observations of such systems using H$\alpha$ emission and polarimetric measurements across the optical and UV appear to be well explained by a disk of material around the white dwarf, and that the resulting accretion onto the white dwarf does well in predicting the observed X-ray fluxes.\\ 

\begin{figure*}
    \centering
    \includegraphics[width=\linewidth]{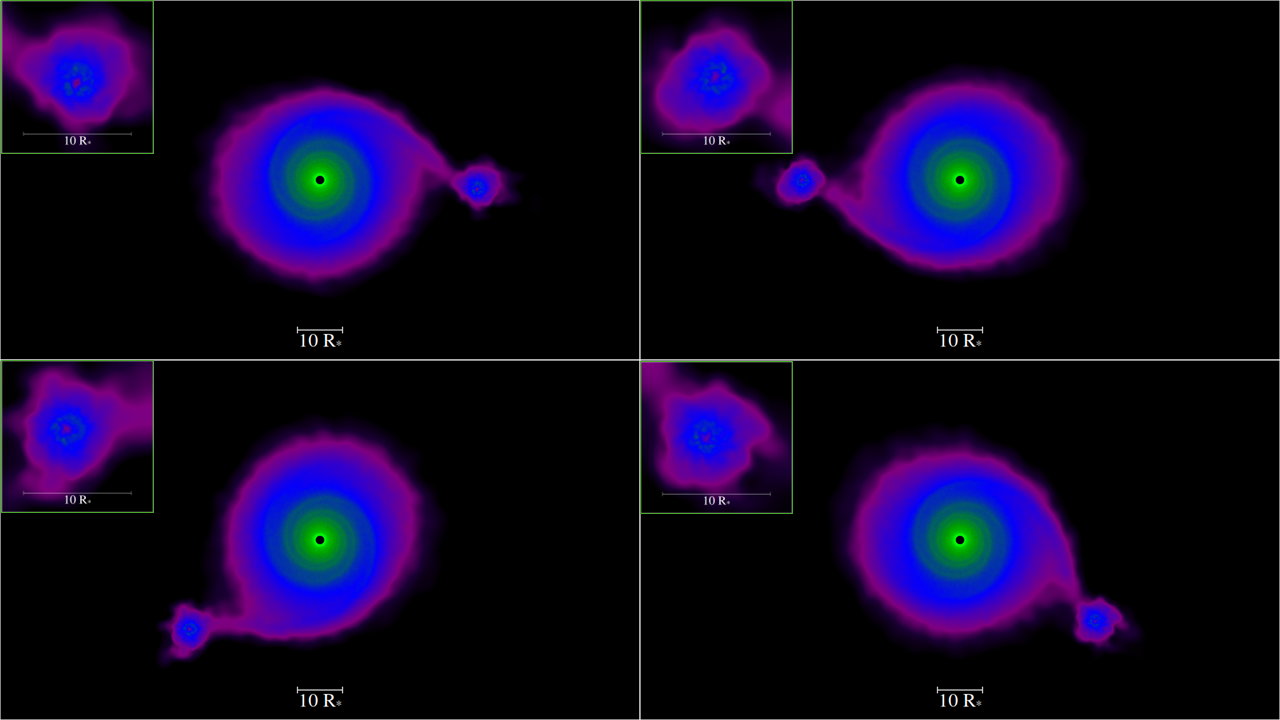}
    \caption{Top down view of $\gamma$ Cas from simulations of Rast et al. (2025, this collection) after quasi-steady state has been reached at 75 orbital periods at four different orbital phases. The insets show enlarged views of the circumsecondary region.}
    \label{fig:Fig2}
\end{figure*}

\item{\bf Angular momentum transport in Be binaries}\\
See \citet{qui25}.

\smallskip
Angular momentum transport is a fundamental process shaping the structure, evolution, and lifespans of stars and disks, especially in the case of Be stars.  These objects provide a somewhat unique and clean environment for studying the impact of viscous angular momentum transport because of their rapid rotation, observable decretion disks, and likely absence of strong magnetic fields.  Using a framework of models that address angular momentum loss in Be stars as a function of orbital separations and companion masses, insight can be gained into the role of angular momentum transport in either direction and can intimately affect the ability of the star to regulate stellar surface rotation.  Predictions can be made about observations that can verify models and constrain stellar parameters using observations in H$\alpha$ line profiles, V-band polarization, and especially UV polarization.\\

\item{\bf Stellar and Interstellar Contributions to the Measured Polarization of Hot Stars}\\
See \citet{ign25}.

\smallskip
Linear polarimetry of unresolved stars is a powerful method for discerning or constraining the geometry of a source and its environment. Since spherical sources produce no net polarization, a general challenge to the interpretation of intrinsic stellar polarization is the contribution to the signal from interstellar polarization (ISP).  Specific contributions to stellar polarization can come from Thomson scattering, the environment of rapidly rotating stars, their optically thick winds, and the dynamics and environment of interacting binaries.  The wavelength-dependent effects of ISP across these scenarios are studied in both the ultraviolet and optical bands.  The results of these comparisons reveal that the impact of ultraviolet spectropolarimetry across the key diagnostic lines can be used to provide key discriminating measurements that separate the effects.\\

\item{\bf Use of the Ohman Effect to Reveal the details of Rotating hot massive stars using linear spectropolarimetry} \\
See \citet{har25b}.

\smallskip
The Ohman effect is the predicted variation in linear polarization across a rotationally broadened absorption line, due to the interaction of that line with the spatially varying continuum polarization across the face of a strongly scattering photosphere, such as found in hot stars.  Such an effect is especially powerful in the far ultraviolet, where the contributions are maximized, and where the density of spectral lines is the greatest, to produce a tell-tale spectral line shape and polarization signatures in the presence of strong rotation.  The Ohman effect can produce, under such circumstances, polarizations as strong as 0.1\% to 1\% across individual lines for a wide variety of B-type stars, and offers an opportunity to access the unique information encoded in the Ohman effect using moderate-resolution spaceborne spectropolarimetric measurements.\\

\item{\bf Polarization from Rapidly Rotating Massive Stars} \\
See \citet{har25a}.

\smallskip
Stellar rotation has long been recognized as important to the evolution of stars, by virtue of the chemical mixing it can induce and how it interacts with binary mass transfer. Binary interaction and rapid rotation are both common in massive stars and involve processes of angular momentum distribution and transport. An important question is how this angular momentum transport leads to the creation of two important classes of rapidly rotating massive stars, Be stars defined by disk-like emission lines, and Bn stars defined by rotationally-broadened absorption lines. A related question is what limits this rotation places on how conservative the mass transfer can be.  Central to addressing these issues is knowledge of how close to rotational break-up stars can get before they produce a disk. It is possible to calculate diagnostics of this rotational criticality using the continuum polarization arising from a combination of rotational stellar distortion (i.e., oblateness) and redistribution of stellar flux (i.e., gravity darkening and polar brightening).  However, polarization signatures in visible light are weak and often dominated by ISM polarization making progress from the ground very difficult. For early main-sequence and subgiant stars, the 
observed polarization signatures can instead reach a maximum of $\sim$ 1\% at 140 nm for stars rotating at 98\% of critical, when seen edge-on. Rotational rates above 80\% critical result in FUV polarizations of several tenths of a percent, at high inclination. Even at a low inclination of i $= 40^o$, models at 98\% critical show polarization in excess of 0.1\% down to 200 nm. These predicted stable signal strengths indicate that determinations of near-critical rotations in B stars could be achieved with future spectropolarimetric instrumentation that can reach deep into the FUV.\\

\item{\bf The Abundances of Nitrogen and Carbon in the Atmospheres of Classical Be Stars}\\
See \citet{Peters26}.

\smallskip
Increasing evidence suggests that mass transfer in a binary system causes the spin-up and disk formation in many Be stars. Given the rapid rotation of the Be star, it is also expected that products of the CNO cycle will be circulated to the photosphere. In this study of eight near pole-on Be stars, mostly all the program stars are found to be reduced in carbon as expected yet there is no corresponding increase in nitrogen. This means that our understanding of the evolution of these systems is too simplistic and incomplete. The authors suggest that the evolutionary timing of the mass transfer from the evolved companion, and if mass transfer occurs more than once, directly impacts the end products. They hypothesize that the lack of excess N could be due to N being converted into O (and perhaps on to Ne) by fusion with He in particular regions of a highly evolved mass loser nearing the end of mass transfer. This work has great potential for the understanding of the spin-up of the Be star and the history of the stripped companion. 
What is needed are observations and analysis of the abundances of C, N, O and Ne for a fulsome understanding of the evolution of these systems. The requisite C, N and O lines are best accessed in the ultraviolet.
\\

\end{itemize}

{\bf Final Remarks} - This collection of papers offers a unique discussion of physical factors that are driven by rapid rotation and whose influence directly impact the evolution and end-of-life state of massive stars.  They are presented together to give the reader a perspective that only the ensemble can provide instead of a single paper. We hope that we are successful in our goal of shedding light on the scope and outcome of this important facet of massive star physics.

\textit{Sincerely the editors}, Paul Scowen, Carol Jones, Ren\'e Oudmaijer, Jamie Lomax, and Jeremy Drake.

\bibliography{prefix}

\end{document}